\RequirePackage{filecontents}

\RequirePackage{snapshot}
\documentclass[aps,pra,reprint,showpacs,secnumarabic,amsmath,amssymb]{preprint}

\renewcommand\documentclass[2][]{}
\let\origparagraph=\paragraph
\AtBeginDocument{%
  \let\paragraph=\origparagraph
}
\makeatother
\documentclass[%
aps,
prd,
twocolumn,
groupedaddress,
floatfix,
amsmath,
amssymb,
longbibliography,
10pt
]{revtex4-1}

\usepackage[utf8]{inputenc}
\usepackage[T1]{fontenc}
\usepackage{amsmath}
\usepackage{txfonts}
\usepackage{tgtermes}
\usepackage[altg]{qtxmath}
\usepackage{braket}
\usepackage{graphicx}
\graphicspath{{figures/}}
\usepackage{color}
\usepackage[english]{babel}
\usepackage{microtype}
\usepackage[breaklinks=true,
pdfauthor={Q.~Z. Lv and Heiko Bauke},
pdftitle={Time- and space-resolved selective multipair creation}]{hyperref}

\newcommand*{\rrangle}{\rangle\!\rangle}

\def\kket#1{\mathinner{\|{#1}\rrangle}}

{\catcode`\|=\active
  \xdef\BBraket{\protect\expandafter\noexpand\csname BBraket \endcsname}
  \expandafter\gdef\csname BBraket \endcsname#1{\begingroup
     \ifx\SavedDoubleVert\relax
       \let\SavedDoubleVert\|\let\|\BraDoubleVert
     \fi
     \mathcode`\|32768\let|\BraVert
     \left\langle\!\left\langle{#1}\right\rangle\!\right\rangle\endgroup}
}

\renewcommand*{\vec}[1]{\boldsymbol{#1}}
\newcommand*{\D}{\mathrm{d}}
\newcommand*{\I}{\mathrm{i}}

\newcommand*{\op}[1]{{\hat#1}}
\newcommand*{\trans}[1]{#1{}^{\mathsf{T}}}
\newcommand*{\htrans}[1]{#1{}^{\mathsf{\dagger}}}

\newcommand*{\commut}[2]{\mathchoice{\left[#1,#2\right]}{[#1,#2]}{[#1,#2]}{[#1,#2]}}
\newcommand*{\acommut}[2]{\mathchoice{\left\{#1,#2\right\}}{\{#1,#2\}}{\{#1,#2\}}{\{#1,#2\}}}
\DeclareMathOperator*{\V}{|}

\begin{document}

\title{Time- and space-resolved selective multipair creation}
\author{Q.~Z. Lv}
\email{qingzheng.lyu@mpi-hd.mpg.de}
\affiliation{Max-Planck-Institut f{\"u}r Kernphysik, Saupfercheckweg 1, 69117 Heidelberg, Germany}

\author{Heiko Bauke}
\email{heiko.bauke@mpi-hd.mpg.de}
\affiliation{Max-Planck-Institut f{\"u}r Kernphysik, Saupfercheckweg 1, 69117 Heidelberg, Germany}

\date{\today}

\begin{abstract}
  The simultaneous creation of multiple electron-positron pairs by
  localized strong electric fields is studied by utilizing a time- and
  space-resolved quantum field theory approach.  It is demonstrated that
  the number of simultaneously created pairs equals the number of the
  potential's supercritical quasibound states in the Dirac sea.  This
  means it can be controlled by tuning the potential parameters.
  Furthermore, the energy of the created particles corresponds to the
  energy of the supercritical quasibound states.  The simultaneously
  created electrons and positrons are statistically correlated, which is
  reflected in the spatial distribution and the momentum distribution of
  these particles and antiparticles.
\end{abstract}

\pacs{34.50.Rk, 03.65.-w, 42.50.-p}

\maketitle

\section{Introduction}
\label{sec:intro}

\enlargethispage{3ex}%
The vacuum is the lowest energy eigenstate of the field-free quantum
field Hamiltonian.  In the presence of a strong external field this
state, however, may become instable.  This leads to the spontaneous
emission of electron-positron pairs, which is one of the most striking
predictions of the Dirac equation in its quantum field theoretical
formulation \cite{Di-Piazza:Muller:etal:2012:Extremely_high-intensity}.
While early predictions of this possibility date back to Heisenberg and
Euler \cite{Heisenberg:Euler:1936:Folgerungen_aus}, Sauter
\cite{Sauter:1931:Uber_Verhalten} and others in the early part of the
past century, the first calculation of the pair-production rate based on
a nonperturbative approach was accomplished by Schwinger
\cite{Schwinger:1951:On_Gauge} in the early 1950s.

Using Schwinger's formula, one finds that a sizable pair-creation rate
requires an electric field of the strength
$E_\mathrm{cr} = m_\mathrm{e} c^2/e = 1.3 \times 10^{18}\,\mathrm{V/m}$,
which is very difficult to produce in the laboratory.  Here
$m_\mathrm{e}$, $e$, and $c$ denote the electron mass, the elementary
charge, and the speed of light.  Further studies
\cite{Hansen:Ravndal:1981:Kleins_Paradox,Holstein:1998:Kleins_paradox}
extended Schwinger’s pioneering work to calculate the long-time
pair-creation behavior for spatially inhomogeneous electric fields.
Several investigations involving the combination of different static
electric, magnetic, and time-dependent laser fields
\cite{Schutzhold:Gies:etal:2008:Dynamically_Assisted,Ruf:Mocken:etal:2009:Pair_Production,Bulanov:Mur:etal:2010:Multiple_Colliding,Wollert:Bauke:etal:2015:Spin_polarized}
suggest that pair creation may be realized below the Schwinger critical
field strength $E_\mathrm{cr}$.

In theoretical terms, the breakdown of the vacuum is associated with a
depopulation of states in the initially filled negative-energy Dirac
sea.  In Ref.~\cite{Muller:Peitz:etal:1972:Solution_of}, it was
suggested that quasibound states that are embedded in the
negative-energy continuum may be solely responsible for pair creation.
If the charge of a combined nucleus is so large that the energies of the
lowest lying bound states drift below $-m_\mathrm{e} c^2$, these states
can dive into the negative energy continuum and trigger pair creation.
Recently, several works
\cite{Jiang:Lv:etal:2013:Enhancement_of,Fillion-Gourdeau:Lorin:etal:2013:Resonantly_Enhanced,Tang:Xie:etal:2013:Electron-positron_pair}
have also argued that discrete states can act as a transfer channel for
population between the positive-energy and negative-energy states and
thus enhance the creation rate.

In fact, starting in the early 1980s heavy ion collision experiments
\cite{Cowan:Backe:etal:1986:Observation_of,Ahmad:Austin:etal:1997:Search_for}
were performed with the hope that the combined Coulomb field of two
colliding nuclei would be sufficient to break down the vacuum
\cite{Greiner:Muller:Rafelski:1985:Quantum_Electrodynamics} and to
produce electron-positron pairs.  However, the observed positrons can
also be caused by the presence of the unavoidable transitions associated
with the internal nuclear structure and not triggered by the Coulomb
field alone.  There is also the prospect that future focused laser
pulses could provide sufficiently large fields to trigger the purely
spontaneous creation of particle pairs from the vacuum.

In recent years, quantum control has become a mature and active research
field which deals with the active manipulation of physical processes on
the quantum level \cite{Brif:Chakrabarti:etal:2010:Control_of}.  Usually
sophisticated schemes are required to drive a quantum system into a
specific desired final state.  It is well known that tuning the
parameters of the binding potential can be utilized to control the
\emph{average} number of created pairs
\cite{Krekora:Cooley:etal:2005:Creation_Dynamics,Lv:Liu:etal:2013:Noncompeting_Channel}.
However, it remains unknown to which field states these pairs actually
belong.  For example, an expectation value for the number of created
pairs of one can correspond to a system that is definitely in a
single-pair state.  The same count can, however, also characterize a
different state that is an equal-weight superposition of the vacuum and
a two-pairs state.  Furthermore, to the best of our knowledge there is
also no method to control a quantum system such that it evolves into a
certain field state during the pair-creation dynamics.  Both issues will
be addressed in this article, which focuses on the pair production
caused by a strong external binding potential.

In this contribution we employ a quantum field theoretical description
of the pair-creation dynamics to study the creation of single-pair
states, two-pair states, and so on.  The numerical time-dependent
solution of the corresponding theoretical equations on a space-time grid
will give us deeper insight than the standard S-matrix approach, which
can only represent the system's asymptotic behavior.  For example, it
allows one to determine the space-resolved densities of multipair states
with a given number of electron-positron pairs.  We will demonstrate
that a strong localized binding potential can be tuned to create
selectively multipair states with a specific number of electrons and
positrons, e.\,g., single-pair states or two-pair states.

This paper is organized as follows.  In order to render the presentation
self-contained, Sec.~\ref{sec:theo} describes the theoretical framework
of numerical time-dependent quantum field theory, which allows us to
investigate the field states as well as the pair-creation dynamics via
arbitrary external force fields.  In Sec.~\ref{sec:pacrea}, we discuss
the pair-creation process for fermionic systems induced by a localized
binding potential and investigate the spatial and temporal signatures of
the final field states.  In Sec.~\ref{sec:Summa}, we give a brief
summary.

\section{Theoretical description of the pair-creation dynamics}
\label{sec:theo}

The creation of particle-antiparticle pairs can be viewed as the vacuum
turns into different field states in the Fock space, i.\,e., the vacuum
state, single-pair states, two-pair states etc.  All these states are
the eigenstates of the field-free quantum field Hamiltonian and can be
used as a basis to span the Hilbert space.  Starting from the vacuum,
the final state of the system is a superposition of several states in
this basis.  As illustrated below, this quantum field state can be
determined by solving the time-dependent Dirac equation for all basis
vectors of the corresponding single-particle Hilbert space
\cite{Gitman:1977:Processes_of,Frolov:Gitman:1978:Density_matrix,Fradkin:etal:1991:Quantum_Electrodynamics,Wollert:Bauke:etal:2016:Multi-pair_states}.
The number of field states that contain a specific count of particles
and antiparticles with different quantum numbers grows exponentially
with the amount of particles.  Consequently, it is very challenging both
analytically as well as numerically to predict the probability to which
particular states will be finally occupied.

During electron-positron pair creation by strong electromagnetic fields,
the initial vacuum state $\kket{\mathrm{vac}}$ evolves into a general
Fock state $\kket{\Omega(t)}$, which is a superposition of the vacuum
state and several multipair states, each containing a specific number of
pairs of particles and antiparticles.  These multipair states are
characterized by a set of various quantum numbers, e.\,g., the kinematic
momentum~$\vec p$ and the spin~$s=\pm\hbar/2$.  In the Schr{\"o}dinger
picture, the quantum field state consisting of particle-antiparticle
multipairs can be written as
\begin{widetext}
  \begin{equation}
    \label{eqn:fieldstate}
    \begin{split}
      \kket{\Omega(t)} = {} & c_0(t)\kket{\mathrm{vac}} + \\
      & c_{1,1}(t)\kket{e^+_{\vec p_1, s_1}e^-_{\vec p'_1, s'_1}} +
      c_{1,2}(t)\kket{e^+_{\vec p_1, s_1}e^-_{\vec p'_2, s'_2}} +
      c_{1,3}(t)\kket{e^+_{\vec p_2, s_2}e^-_{\vec p'_1, s'_1}} + \cdots \\
      & c_{2,1}(t)\kket{e^+_{\vec p_1, s_1}e^+_{\vec p_2, s_2}e^-_{\vec p'_1, s'_1}e^-_{\vec p'_2, s'_2}}+
      c_{2,2}(t)\kket{e^+_{\vec p_1, s_1}e^+_{\vec p_3, s_3}e^-_{\vec p'_1, s'_1}e^-_{\vec p'_2, s'_2}}+
      c_{2,3}(t)\kket{e^+_{\vec p_1, s_1}e^+_{\vec p_2, s_2}e^-_{\vec p'_1, s'_1}e^-_{\vec p'_3, s'_3}}+
      \cdots \\
      &
      c_{3,1}(t)\kket{e^+_{\vec p_1, s_1}e^+_{\vec p_2, s_2}e^+_{\vec p_3, s_3}e^-_{\vec p'_1, s'_1}e^-_{\vec p'_2, s'_2}e^-_{\vec p'_3, s'_3}}+
      c_{3,2}(t)\kket{e^+_{\vec p_1, s_1}e^+_{\vec p_2, s_2}e^+_{\vec p_4, s_4}e^-_{\vec p'_1, s'_1}e^-_{\vec p'_2, s'_2}e^-_{\vec p'_3, s'_3}}+% \\
      % & \hspace{50ex}
      % c_{3,3}(t)\kket{e^+_{\vec p_1, s_1}e^+_{\vec p_2, s_2}e^+_{\vec p_3, s_3}e^-_{\vec p'_1, s'_1}e^-_{\vec p'_2, s'_2}e^-_{\vec p'_4, s'_4}}+
      \cdots\,,
    \end{split}
  \end{equation}
\end{widetext}
where $\kket{\mathrm{vac}}$ denotes the vacuum state,
$\kket{e^+_{\vec p_1, s_1}e^-_{\vec p'_1, s'_1}}$ is a single-pair state
with an electron ($e^+$) and a positron ($e^-$) \footnote{Note that
  upper indices $+/-$ refer to the sign of the particles' energy as
  within the single-particle picture of the Dirac theory, not to the
  particles' charge.},
$\kket{e^+_{\vec p_1, s_1}e^+_{\vec p_2, s_2}e^-_{\vec p'_1,
    s'_1}e^-_{\vec p'_2, s'_2}}$ refers a two-pair state and so on.  A
superposition of multipair states with all of the same number of
particle-antiparticle pairs is called a ``number state'' in the
following.  Mathematically, a multipair state with $n$ pairs is created
from the vacuum state by applying the creation operators
$\htrans{\op a_{\vec p, s}^+\!}$ and $\htrans{\op a_{\vec p, s}^-\!}$
for the particle and the antiparticle, respectively, on it, i.\,e.,
\begin{equation}
  \kket{e^+_{\vec p_1, s_1}\dots e^+_{\vec p_n,s_n}e^-_{\vec p'_1,
      s'_1}\dots e^-_{\vec p'_n, s'_n}}=
  \prod_{i=1}^n
  \htrans{\op a_{\vec p_i,s_i}^+\!}\,
  \htrans{\op a_{\vec p'_i, s'_i}^-\!}\kket{\mathrm{vac}} \,.
\end{equation}
The corresponding annihilation operators will be denoted by $\op a_{\vec
  p, s}^+$ and $\op a_{\vec p, s}^-$.  These fermionic annihilation and
creation operators satisfy the anticommutator relations
\begin{equation}
  \label{eq:creation_comm}
  \acommut{\op a_{\vec p, s}^+}{\htrans{\op a_{\vec p', s'}^+\!}}=
  \acommut{\op a_{\vec p, s}^-}{\htrans{\op a_{\vec p', s'}^-\!}}=
  \delta_{\vec p, \vec p'}\delta_{s, s'}  \,,
\end{equation}
where $\delta_{i, j}$ denotes a Kronecker delta.  The amplitudes of
these states are represented by $c_0(t)$ for vacuum and $c_{i,j}(t)$ for
the $j$th single- or multipair state containing $i$ electrons and
positrons.  As the number of ways to combine particle-antiparticle pairs
into multipair states grows rapidly with the number of pairs it is not
feasible to calculate the amplitudes for all states.  Instead, we define
various observables based on the quantum field operator to characterize
the quantum field state in the following.

The quantum field operator of a fermionic many-particle system can be
expressed as an integral or a sum (in the case of a discretized
Hamiltonian) over the electronic annihilation operators and the
positronic creation operators,
\begin{equation}
  \label{eq:field_operator}
  \op\Psi(\vec r) =
  \sum_{\vec p, s} \op a_{\vec p, s}^+\psi_{\vec p, s}^+(\vec r) +
  \sum_{\vec p, s} \htrans{\op a_{\vec p, s}^-\!}\psi_{\vec p, s}^-(\vec r)\,.
\end{equation}
Here, $\psi_{\vec p, s}^+(\vec r)$ denotes a normalized eigenstate of
the free Dirac equation with positive energy, the momentum eigenvalue
$\vec p$, and the spin $s$, and correspondingly
$\psi_{\vec p, s}^-(\vec r)$ denotes an eigenstate with negative energy.
This means, $\psi_{\vec p, s}^+(\vec r)$ and
$\psi_{\vec p, s}^-(\vec r)$ are eigenfunctions of the
first-quantization Dirac Hamiltonian
\begin{equation}
  \op{H}_\mathrm{D} =
  c \vec\alpha\cdot(\vec{\op p} -q \vec A(\vec r, t))  +
  \beta m_\mathrm{e}c^2 + q\phi(\vec r, t)
\end{equation}
in the special case of a vanishing electric potential $\phi(\vec r, t)$
and a vanishing magnetic vector potential $\vec A(\vec r, t)$.  Here, we
also introduced the momentum operator $\vec{\op p}$, the electron's
charge $q=-e$, as well as the Dirac matrices
$\vec{\alpha}=\trans{(\alpha_1, \alpha_2, \alpha_3)}$ and $\beta$.
Adopting the Heisenberg picture, the operators $\op\Psi(\vec r)$ and
$\op a_{\vec p, s}^\pm$ become time dependent.  The time-dependent
field operator $\op\Psi(\vec r, t)$ is given in terms of time-dependent
creation and annihilation operators by
\begin{equation}
  \label{eq:field_operator_2}
  \op\Psi(\vec r, t) =
  \sum_{\vec p, s} \op a_{\vec p, s}^+(t)\psi_{\vec p, s}^+(\vec r) +
  \sum_{\vec p, s} \htrans{\op a_{\vec p, s}^-(t)}\psi_{\vec p, s}^-(\vec r)\,.
\end{equation}

Stripping the antiparticle part from the quantum field operator
\eqref{eq:field_operator_2}, we define the operator
\cite{Schweber:2005:QFT}
\begin{equation}
  \label{eq:field_operator_particle}
  \op\Psi^+(\vec r, t) =
  \sum_{\vec p, s} \op a_{\vec p, s}^+(t)\psi_{\vec p, s}^+(\vec r) \,.
\end{equation}
With this definition operators representing various physical observables
can be established, e.\,g., the generalized particle-number density
operators
\begin{multline}
  \label{eq:op_N_t}
  \op N_n(t) =
  \frac{1}{n!}
  \htrans{\op\Psi^+(\vec r_1, t)}\cdots\htrans{\op\Psi^+(\vec r_n, t)}
  \op\Psi^+(\vec r_n, t)\cdots\op\Psi^+(\vec r_1, t)
\end{multline}
can be introduced for $n=1,2,\dots$.  The average density
$\varrho(\vec r_1,\dots,\vec r_n, t)$ of finding simultaneously
particles at the positions $\vec r_1$, \dots, $\vec r_n$ at time $t$ if
the system was initially in the quantum field state $\kket{\Omega}$ is
given by
\begin{equation}
  \label{eq:rho_n}
  \varrho(\vec r_1,\dots,\vec r_n, t) =
  \BBraket{\Omega\| \op N_n(t) \|\Omega}\,.
\end{equation}
Integrating over the whole space leads to the expectation values of the
generalized particle number operators,
\begin{equation}
  \label{eq:N_n_t}
  N_n(t) =
  \int\cdots\int \varrho(\vec r_1,\dots,\vec r_n, t) \,\D^3 r_1\cdots\,\D^3 r_n \,.
\end{equation}
The quantity $N_1(t)$ denotes the average number of particles at time
$t$.  In general, $N_n(t)$ is the average number of $n$-tuples of
particles present at time $t$.  These $n$-tuples may originate from a
number state with $n$~particles or from some state containing $m>n$
particles.  In the special case that the initial state $\kket{\Omega}$
has evolved into a single number state of $m$ pairs then
$N_n(t)=\binom{m}{n}$ because there are $\binom{m}{n}$ ways to pick $n$
particles from a set of $m$ particles.  This can also be confirmed by an
explicit calculation of $N_n(t)$ for some $m$-particle state.
Consequently, we can also write
\begin{equation}
  \label{eq:N_n_t_C_m_t}
  N_n(t) = \sum_{m=n}^\infty \binom{m}{n}C_m(t) \,,
\end{equation}
where $C_m(t)$ denotes the total probability that $\kket{\Omega}$ has
evolved into some number state with $m$ pairs.  This means with respect
to the expansion coefficients $c_{n,m}$ of a quantum field state in the
Schr{\"o}dinger picture as in Eq.~\eqref{eqn:fieldstate}
\begin{equation}
  C_n(t) = \sum_{m} |c_{n,m}(t)|^2 \,.
\end{equation}
The set of linear equations~\eqref{eq:N_n_t_C_m_t} can be inverted to
yield
\cite{Cheng:Su:etal:2009:Creation_of,Su:Li:etal:2012:Non-perturbative_approach}
\begin{equation}
  \label{eq:C_n_t_N_m_t}
  C_n(t) = \sum_{m=n}^\infty (-1)^{m+n}\binom{m}{n}N_m(t) \,.
\end{equation}
The $C_n(t)$ characterize how many particle-antiparticle pairs are
created preferably by the external strong electromagnetic fields.  In
analogy to the $n$-pair probability $C_n(t)$ as given in
Eq.~\eqref{eq:C_n_t_N_m_t}, we find the $n$-pair probability density
\begin{multline}
  \rho(\vec r_1,\dots,\vec r_n, t) = \\
  \sum_{m=n}^\infty (-1)^{m+n}\binom{m}{n}
  \int\cdots\int\varrho(\vec r_1,\dots,\vec r_m, t) \,\D^3 r_{n+1}\cdots\,\D^3 r_m \,.
\end{multline}
Here, position variables $\vec r_{n+1}$, \dots, $\vec r_m$ are
integrated out.  The choice of the integration variables, however, is of
no relevance as the density $\rho(\vec r_1,\dots,\vec r_m, t)$ is
invariant under permutation of the position variables.  Note that if the
quantum field state is a superposition of $n$-pair states only, i.\,e.,
$C_n(t)=1$, then $\rho(\vec r_1,\dots,\vec r_n, t)=\varrho(\vec
r_1,\dots,\vec r_n, t)$.

To determine the densities defined in Eq.~\eqref{eq:rho_n} and the
expectation values \eqref{eq:N_n_t} and \eqref{eq:C_n_t_N_m_t}, which
are of main interest here, we need to solve the time dependence of the
quantum field operator.  The time-dependent quantum field operators as
well as the creation and the annihilation operators fulfill the
Heisenberg equations of motion
\begin{equation}
  \label{eq:Heisenberg_Psi}
  \I\hbar \frac{\partial \op \Psi(\vec r, t)}{\partial t} =
  \commut{\op\Psi(\vec r, t)}{\op H}
\end{equation}
and
\begin{equation}
  \I\hbar \frac{\partial \op a_{\vec p, s}^\pm(t)}{\partial t} =
  \commut{\op a_{\vec p, s}^\pm(t)}{\op H} \,,
\end{equation}
\vspace*{1ex plus 1ex}

\noindent
respectively, with the Hamilton operator
\begin{equation}
  \label{eq:op_H}
  \op H = \int
  \htrans{\op\Psi}(\vec r, t)\op{H}_\mathrm{D}\op\Psi(\vec r, t)
  \,\D^3 r \,.
\end{equation}
The equation of motion \eqref{eq:Heisenberg_Psi} can be further
simplified via Eq.~\eqref{eq:op_H} to the Schr{\"o}dinger-like equation
\cite{Geiner:Reinhardt:1996:Field_quantization}
\begin{equation}
  \label{eq:Heisenberg_Psi_bar}
  \I\hbar \frac{\partial \op \Psi(\vec r, t)}{\partial t} =
  \op{H}_\mathrm{D}\op\Psi(\vec r, t) \,.
\end{equation}
Consequently, the time-dependent field operator $\op \Psi(\vec r, t)$
can also be expressed as
\begin{equation}
  \label{eq:field_operator_3}
  \op\Psi(\vec r, t) =
  \sum_{\vec p, s} \op a_{\vec p, s}^+\psi_{\vec p, s}^+(\vec r, t) +
  \sum_{\vec p, s} \htrans{\op a_{\vec p, s}^-}\psi_{\vec p, s}^-(\vec r, t)\,,
\end{equation}
where the functions $\psi_{\vec p, s}^+(\vec r, t)$ and
$\psi_{\vec p, s}^-(\vec r, t)$ denote the solutions of the
time-dependent Dirac equation with $\psi_{\vec p, s}^+(\vec r)$ and
$\psi_{\vec p, s}^-(\vec r)$, respectively, as initial conditions at
time $t=0$ and with $\op a_{\vec p, s}^+=\op a_{\vec p, s}^+(0)$ and
$\htrans{\op a_{\vec p, s}^-}= \htrans{\op a_{\vec p, s}^-}(0)$.
Equating Eqs.~\eqref{eq:field_operator_2} and
\eqref{eq:field_operator_3}, we can solve for $\op a_{\vec p, s}^+(t)$
and $\htrans{\op a_{\vec p, s}^-} (t)$ and finally find
\cite{Krekora:Su:etal:2004:Klein_Paradox,Cheng:Su:etal:2010:Introductory_review}
\begin{equation}
  \op a_{\vec p, s}^+(t) =
  \sum_{\vec p', s'}
  G(\sideset{^+}{_{+}}\V)_{\vec p, s; \vec p', s'}\op a_{\vec p', s'}^+ +
  G(\sideset{^+}{_{-}}\V)_{\vec p, s; \vec p', s'}\htrans{\op a_{\vec p', s'}^-}
\end{equation}
and
\begin{equation}
  \htrans{\op a_{\vec p, s}^-} (t) =
  \sum_{\vec p', s'}
  G(\sideset{^-}{_{+}}\V)_{\vec p, s; \vec p', s'}\htrans{\op a_{\vec p', s'}^-} +
  G(\sideset{^-}{_{-}}\V)_{\vec p, s; \vec p', s'}\op a_{\vec p', s'}^+
\end{equation}
with the transition amplitudes
\begin{equation}
  G(\sideset{^\nu}{_{\nu'}}\V)_{\vec p, s; \vec p', s'} =
  \braket{\psi_{\vec p, s}^\nu(\vec r)|\psi_{\vec p', s'}^{\nu'}(\vec r, t)} \,.
\end{equation}

Assuming that initially the pure vacuum state
$\kket{\Omega}=\kket{\mathrm{vac}}$ is given, the density
$\varrho(\vec r_1,\dots,\vec r_n, t)$ can be written by employing
Eq.~\eqref{eq:creation_comm} and
$\op{a}^\pm_{\vec p, s}\kket{\mathrm{vac}}=0$ and introducing the
Hermitian matrix
\begin{equation}
  S_{\vec p, s; \vec p', s'}(t) = \sum_{\vec p'', s''}
  G(\sideset{^+}{_{-}}\V)_{\vec p, s; \vec p'', s''}^*
  G(\sideset{^+}{_{-}}\V)_{\vec p', s'; \vec p'', s''}
\end{equation}
as
\begin{widetext}
  \begin{multline}
    \label{eq:rho_n_expl}
    \varrho(\vec r_1,\dots,\vec r_n, t) =
    \frac{1}{n!}
    \sum_{\substack{\vec p_1, \vec p_2, \dots, \vec p_n \\
        s_1, s_2, \dots, s_n\\
        \vec p'_1, \vec p'_2, \dots, \vec p'_n \\
        s'_1, s'_2, \dots, s'_n}}
    \left(
      \sum_{(i_1, i_2, \dots, i_n)\in P_n} \sigma_{i_1, i_2, \dots, i_n}
      S_{\vec p_1, s_1; \vec p'_{i_1}, s'_{i_1}}(t)S_{\vec p_2, s_2; \vec p'_{i_2}, s'_{i_2}}(t)\dots S_{\vec p_n, s_n; \vec p'_{i_n}, s'_{i_n}}(t)
    \right)\\[-4ex]
    \times \htrans{\psi_{\vec p_1, s_1}^+(\vec r_1)}\htrans{\psi_{\vec p_2,
        s_2}^+(\vec r_2)}\dots\htrans{\psi_{\vec p_n, s_n}^+(\vec r_n)}
    \psi_{\vec p'_n, s'_n}^+(\vec r_n)\dots\psi_{\vec p'_2, s'_2}^+(\vec
    r_2)\psi_{\vec p'_1, s'_1}^+(\vec r_1) \,,
  \end{multline}
  where the innermost sum runs over all permutations $(i_1, i_2, \dots,
  i_n)$ of the set of the natural numbers $1$ to $n$ and $\sigma_{i_1,
    i_2, \dots, i_n}$ denotes the permutation's sign.  Integrating
  \eqref{eq:rho_n_expl} over the whole space, one gets
  \begin{equation}
    \label{eq:N_n_expl}
    N_n(t) = \frac{1}{n!}
    \sum_{\substack{\vec p_1, \vec p_2, \dots, \vec p_n\\s_1, s_2, \dots, s_n}}
    %\left(
      \sum_{(i_1, i_2, \dots, i_n)\in P_n} \sigma_{i_1, i_2, \dots, i_n}
      S_{\vec p_1, s_1; \vec p_{i_1}, s_{i_1}}(t)
      S_{\vec p_2, s_2; \vec p_{i_2}, s_{i_2}}(t)\dots
      S_{\vec p_n, s_n; \vec p_{i_n}, s_{i_n}}(t)
      % \right)
      \,.
  \end{equation}
  Note that the innermost sums in \eqref{eq:rho_n_expl} and
  \eqref{eq:N_n_expl} represent a determinant.  The outer sum over
  momenta and spins in \eqref{eq:rho_n_expl} contains for each summand
  also its complex conjugate, thus $\varrho(\vec r_1,\dots,\vec r_n, t)$
  is real valued.  Furthermore, the outer sum over momenta and spins in
  \eqref{eq:N_n_expl} runs over determinants of Hermitian matrices, thus
  also $N_n(t)$ is real valued. For $n=1$ and $n=2$,
  Eqs.~\eqref{eq:rho_n_expl} and \eqref{eq:N_n_expl} simplify to
  \begin{subequations}
    \begin{align}
      \varrho(\vec r_1, t) & =
      \sum_{\substack{\vec p_1, s_1\\\vec p'_1,s'_1}} S_{\vec p_1, s_1;\vec p'_1, s'_1}(t)
      \htrans{\psi_{\vec p_1, s_1}^+(\vec r_1)}\psi_{\vec p'_1, s'_1}^+(\vec r_1) \,, \\
      \varrho(\vec r_1, \vec r_2, t) & =
      \frac{1}{2}
      \sum_{\substack{\vec p_1, \vec p_2, s_1, s_2\\\vec p'_1, \vec p'_2, s'_1, s'_2}} \left(
        S_{\vec p_1, s_1; \vec p'_1, s'_1}(t)
        S_{\vec p_2, s_2; \vec p'_2, s'_2}(t) -
        S_{\vec p_1, s_1; \vec p'_2, s'_2}(t)
        S_{\vec p_2, s_2; \vec p'_1, s'_1}(t)
      \right)
      \htrans{\psi_{\vec p_1,s_1}^+(\vec r_1)}\htrans{\psi_{\vec p_2,s_2}^+(\vec r_2)}\psi_{\vec p'_2,s'_2}^+(\vec r_2)\psi_{\vec p'_1,s'_1}^+(\vec r_1)
      \,,
    \end{align}
  \end{subequations}
and
\begin{subequations}
  \begin{align}
    N_1(t) & = \sum_{\vec p_1, s_1} S_{\vec p_1, s_1; \vec p_1, s_1}(t) \,, \\
    N_2(t) & = \frac{1}{2}\sum_{\vec p_1, s_1, \vec p_2, s_2} \left(
      S_{\vec p_1, s_1; \vec p_1, s_1}(t)
      S_{\vec p_2, s_2; \vec p_2, s_2}(t) -
      S_{\vec p_1, s_1; \vec p_2, s_2}(t)
      S_{\vec p_2, s_2; \vec p_1, s_1}(t)
    \right) \,.
  \end{align}
\end{subequations}
Similarly to the position-space distribution
$\varrho(\vec r_1,\dots,\vec r_n,t)$ one can introduce a momentum-space
distribution $\chi_n(\vec p_1, \dots, \vec p_n, t)$ of the created
particles, which reads
\begin{equation}
  \chi_n(\vec p_1, \dots, \vec p_n, t) = \frac{1}{n!}
  \sum_{\substack{s_1, s_2, \dots, s_n}}
  \sum_{(i_1, i_2, \dots, i_n)\in P_n} \sigma_{i_1, i_2, \dots, i_n}
  S_{\vec p_1, s_1; \vec p_{i_1}, s_{i_1}}(t)
  S_{\vec p_2, s_2; \vec p_{i_2}, s_{i_2}}(t)\dots
  S_{\vec p_n, s_n; \vec p_{i_n}, s_{i_n}}(t) \,.
\end{equation}
Accordingly,  the single-particle and double-particle distributions
simplify to
\begin{subequations}
  \begin{align}
    \chi_1(\vec p_1, t)
    & =
      \sum_{\substack{s_1}} S_{\vec p_1, s_1;\vec p_1, s_1}(t)\,, \\
    \chi_2(\vec p_1, \vec p_2, t)
    & =
      \frac{1}{2}
      \sum_{\substack{s_1, s_2}} \left(
      S_{\vec p_1, s_1; \vec p_1, s_1}(t)
      S_{\vec p_2, s_2; \vec p_2, s_2}(t) -
      S_{\vec p_1, s_1; \vec p_2, s_2}(t)
      S_{\vec p_2, s_2; \vec p_1, s_1}(t)
      \right)\,.
  \end{align}
\end{subequations}

Note that densities and expectation values of particle numbers for the
antiparticle part of the created pairs can be derived, too, analogously
to the particle case as outlined above.  For this purpose, one has to
replace the operator~\eqref{eq:field_operator_particle} in the
definition of the particle-number density operator~\eqref{eq:op_N_t} by
\begin{equation}
  \label{eq:field_operator_antiparticle}
  \op\Psi^-(\vec r, t) =
  \sum_{\vec p, s} \htrans{\op a_{\vec p, s}^-(t)}\psi_{\vec p, s}^-(\vec r)\,.
\end{equation}
In particular, the momentum-space distribution
$\chi_n^-(\vec p_1, \dots, \vec p_n, t)$ of the created antiparticles
yields 
  \begin{equation}
    \chi_n^-(\vec p_1, \vec p_2, \dots, \vec p_n, t) = \frac{1}{n!}
    \sum_{\substack{s_1, s_2, \dots, s_n}}
    \sum_{(i_1, i_2, \dots, i_n)\in P_n} \sigma_{i_1, i_2, \dots, i_n}
    S^-_{\vec p_1, s_1; \vec p_{i_1}, s_{i_1}}(t)
    S^-_{\vec p_2, s_2; \vec p_{i_2}, s_{i_2}}(t)\dots
    S^-_{\vec p_n, s_n; \vec p_{i_n}, s_{i_n}}(t) \,,
  \end{equation}
  where the matrix
  $S^-_{\vec p, s; \vec p', s'}(t) = \sum_{\vec p'', s''}G(\sideset{^{-}}{_+}\V)_{\vec p, s;
    \vec p'', s''}^*G(\sideset{^{-}}{_+}\V)_{\vec p', s'; \vec p'',
    s''}$ has been introduced.
  \pagebreak
\end{widetext}

\section{Controlling pair creation triggered by a strong localized
  potential}
\label{sec:pacrea}

\subsection{Energy spectrum of a strong localized potential}

The Coulomb binding potential supports several electronic bound states
and with increasing potential depth these bound states can dive into the
negative-energy continuum
\cite{Greiner:1997:Relativistic_Quantum_Mechanics}.  These embedded
states can be viewed as the supercritical quasibound states, which build
up the connection between positive and negative energy levels and in
this way trigger spontaneous pair creation.  The essential physics of
this process can be represented by a one-dimensional model system, which
we will employ for the remainder of this article.  Note that in one
dimension the Dirac equation reduces to an equation for two-component
wave function with no spin \cite{Thaller:2005:vis_QM_II}.  For numerical
feasibility, we choose a localized scalar potential well of the form
\begin{equation}
  \label{eq:well_pot}
  q\phi(x, t) = V_0\big(S(x+D/2)-S(x-D/2)\big)f(t)
\end{equation}
instead of the long range Coulomb field.  Here the parameter $D$ is
related to the spatial width of the well, which is formed by two
smooth unit-step functions
\begin{equation}
  S(x) =  \frac{1}{2}\left(1 + \tanh \frac x W \right)\,,
\end{equation}
where $W$ is the extent of the associated localized electric fields
\cite{Sauter:1931:Uber_Verhalten}.  The time-dependent function $f(t)$
describes the temporal profile of the external field.  In our
calculation, we employ
\begin{equation}
  \label{eq:turn_on_and_off}
  f(t)=
  \begin{cases}
    \sin^2\frac{\pi(t-\Delta T)}{2\Delta T} &
    \text{for $-\Delta T\le t\le 0$}\,, \\
    1, & 0 \le t \le T \\
    \cos^2\frac{\pi (t-T)}{2\Delta T} &
    \text{for $T \le t \le T+\Delta T$} \,,
  \end{cases}
\end{equation}
where $T$ denotes the period of the flat plateau and $\Delta T$ the
duration of turn-on and turn-off.

\begin{figure}
  \centering
  \includegraphics[scale=0.485]{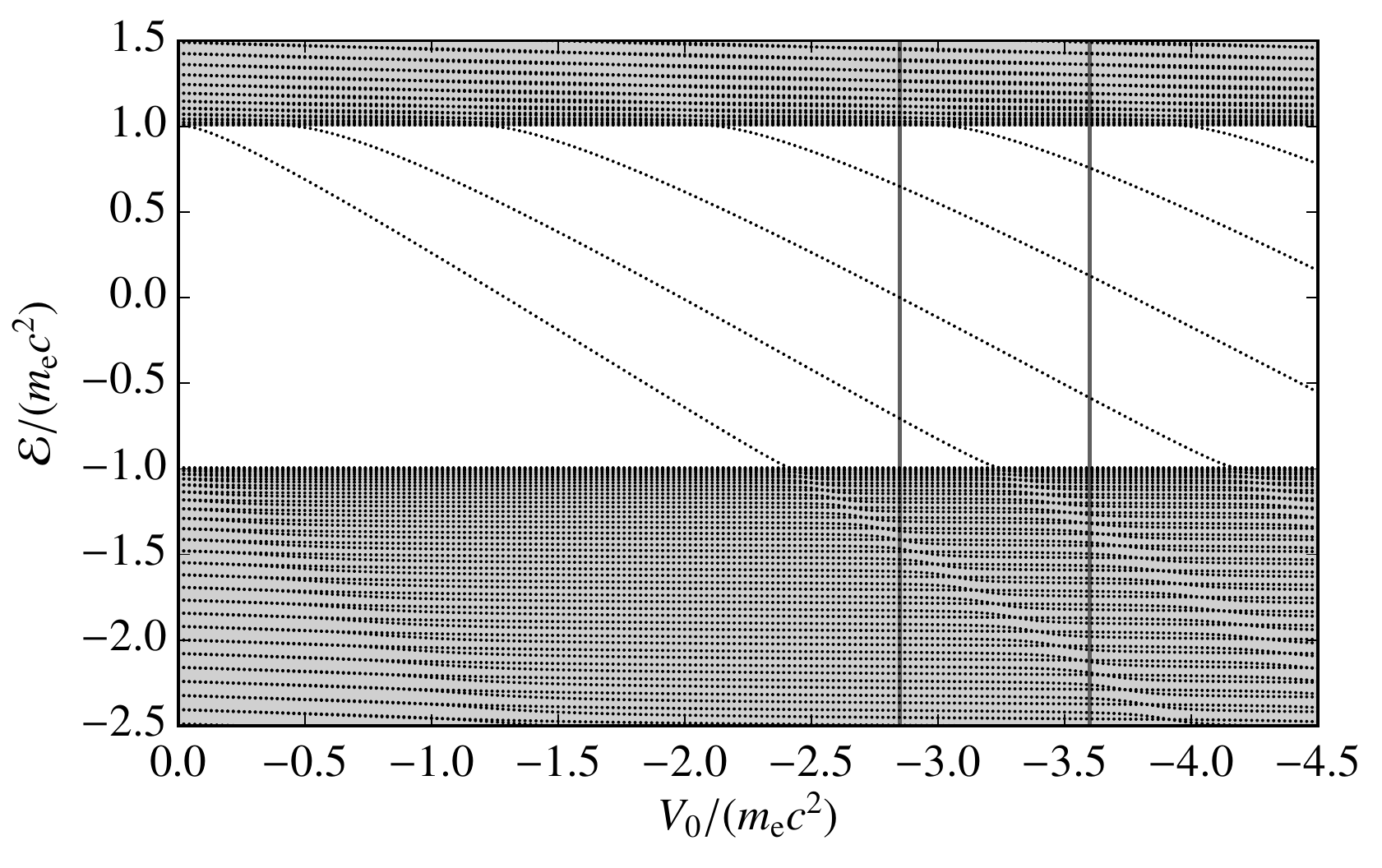}
  \caption{Energy eigenvalues $\mathcal{E}$ of the discretized
    one-dimensional Dirac Hamiltonian with the potential
    \eqref{eq:well_pot} with $W=0.3\lambda_\mathrm{C}$ and
    $D=3.2\lambda_\mathrm{C}$ as a function of the potential strength
    $V_0$ with $\lambda_\mathrm{C}$ denoting the Compton wavelength.
    Gray shaded areas indicate the continuous spectrum of the underlying
    Dirac Hamiltonian $\op H_\mathrm{D}$.  For the discretization of the
    Dirac Hamiltonian a regular spatial grid running from
    $-34.25\lambda_\mathrm{C}$ to $34.25\lambda_\mathrm{C}$ with $512$
    grid points was applied.  The two vertical lines indicate the
    potential strengths that are employed in
    Figs.~\ref{fig:multi_pair_creation_C_V0}(a) and
    \ref{fig:multi_pair_creation_C_V0}(b).}
  \label{fig:energy_spectrum_well}
\end{figure}

The field configuration at the plateau phase $0 \le t \le T$ can support
several electronic bound states and as the potential height $V_0$
increases, the lower bound states can overlap with the negative-energy
continuum.  The resulting degeneracy between the quasibound states and
the negative-energy continuum leads in the case of a discretized Dirac
Hamiltonian, as it is employed in our numerical calculations, to an
increased density of states as shown in
Fig.~\ref{fig:energy_spectrum_well}.  By varying the potential strength
$V_0$ (or its width $W$), we can control the number of supercritical
quasibound states in the negative continuum.  As each number state has
some energy that is at least $2m_\mathrm{e}c^2$ times the number of
pairs, it is commonly believed that single-pair states are preferably
created over states consisting of several pairs.  However, we will show
in the following that if there is more than one supercritical quasibound
state in the Dirac sea, the system will prefer to populate multipair
states rather than single-pair states, which can be strongly suppressed.

\begin{figure}
  \centering
  \includegraphics[scale=0.485]{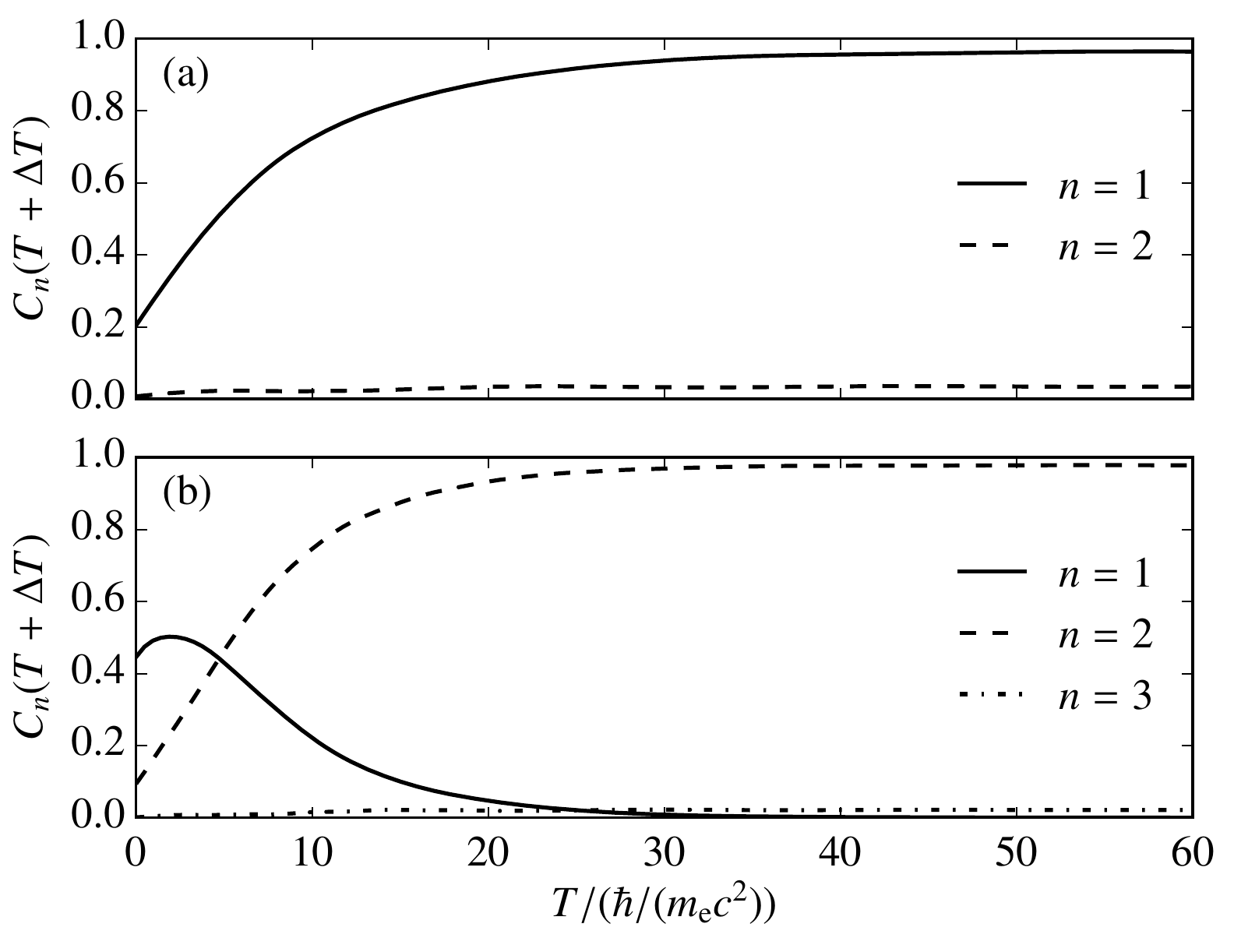}
  \caption{$n$-pair probabilities $C_n$ as a function of the interaction
    time $T$ after the potential has been smoothly turned off with
    $\Delta T=4.7\hbar/(m_\mathrm{e}c^2)$ for $V_0=-2.85m_\mathrm{e}c^2$
    in~(a) and $V_0=-3.6m_\mathrm{e}c^2$ in~(b). Other parameters are as
    in Fig.~\ref{fig:energy_spectrum_well}.}
  \label{fig:multi_pair_creation_C_V0}
\end{figure}

\subsection{Number states in pair creation}

Figure~\ref{fig:multi_pair_creation_C_V0} presents the probability $C_n$
to create a number state of $n$ pairs as a function of the interaction
time $T$ after the potential has been smoothly turned off at time
$T+\Delta T$.  In Fig.~\ref{fig:multi_pair_creation_C_V0}(a) with
$V_0=-2.85m_\mathrm{e}c^2$, the quantum field state evolves into a
superposition of single-pair states as $C_1(T+\Delta T)\approx 1$ for
sufficiently long interaction times.  There is only a small probability
to populate two-pair states. The quantities $C_n(T+\Delta T)$ for $n>2$
are so small that they cannot be distinguished from zero on the scale of
Fig.~\ref{fig:multi_pair_creation_C_V0} and are therefore not shown.  As
further numerical simulations show, the nonzero probability
$C_2(T+\Delta T)$ results from the nonadiabatic turn-on and turn-off and
can be reduced by switching the potential on and off more slowly.
Apparently the creation of sole single-pair states is related to the
fact that exactly one quasibound state in the negative continuum is
present for the chosen potential parameters.  Increasing the depth of
the potential to $V_0=-3.6m_\mathrm{e}c^2$ creates a second quasibound
state as indicated in Fig.~\ref{fig:energy_spectrum_well}.  The
\mbox{$n$-pair} probabilities $C_n$ for this setup are presented in
Fig.~\ref{fig:multi_pair_creation_C_V0}(b).  For short interaction times
we find a superposition of single-, two-, and three-pair states.
Similar to the former case the nonzero probability $C_3(T+\Delta T)$
results from the nonadiabatic turn-on and turn-off.  More interestingly,
the single-pair states present for short interaction times disappear
after some transient interval of interaction times $T$ and for
sufficiently long interaction times the quantum field state is almost
only a superposition of two-pair states, i.\,e.,
$C_2(T+\Delta T)\approx1$.

\begin{figure}
  \centering
  \includegraphics[scale=0.485]{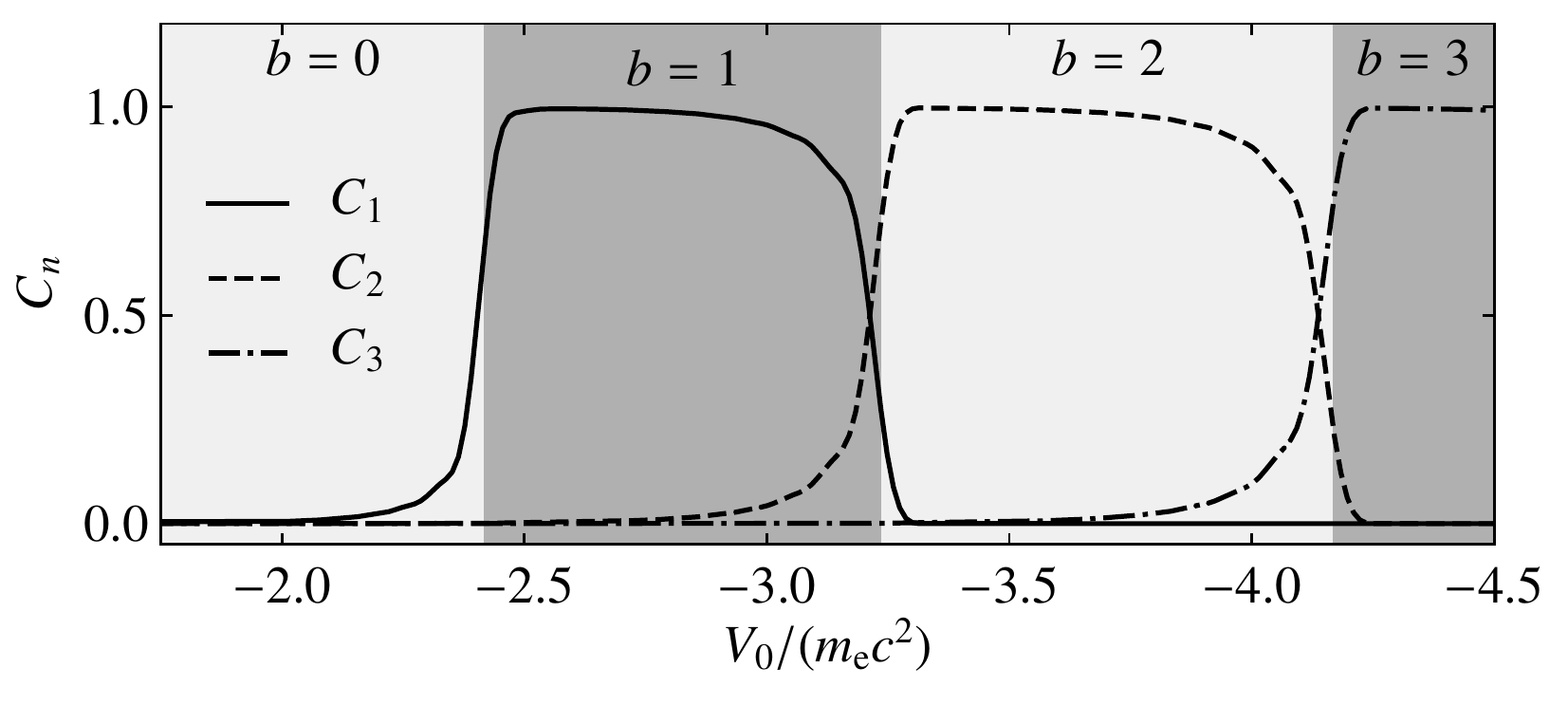}
  \caption{$n$-pair probabilities $C_n$ as a function of the maximal
    potential strength $V_0$ after the potential has been smoothly
    turned off with a total interaction time
    $T=93\hbar/(m_\mathrm{e}c^2)$.  Other parameters are as in
    Fig.~\ref{fig:multi_pair_creation_C_V0}.  The gray shaded areas
    indicate parameter regions with a particular number of supercritical
    quasibound states $b$.}
  \label{fig:multi_pair_creation_V0_Cn}
\end{figure}

Our numerical results indicate that the number of the potential's
quasibound states determines into what kind of number state the quantum
field state evolves for sufficiently long interaction times.  This is
also illustrated in Fig.~\ref{fig:multi_pair_creation_V0_Cn}, where the
probability $C_n$ to create a number state of $n$ pairs is shown as a
function of the maximal potential strength $V_0$ and as a function of
the number of supercritical quasibound states $b$.  For potential
parameters with $b$ supercritical quasibound states we find
$C_b\approx 1$, except for parameters close to a change of the value of
$b$.  This means one can control into which particle-number state the
system will eventually evolve just by changing the number of
supercritical quasibound states in the system.  At least in principle,
this number can be easily adjusted by varying the strength and the shape
of the supercritical electric potential as the only two relevant
parameters.

\begin{figure}%
  \centering
  \includegraphics[scale=0.485]{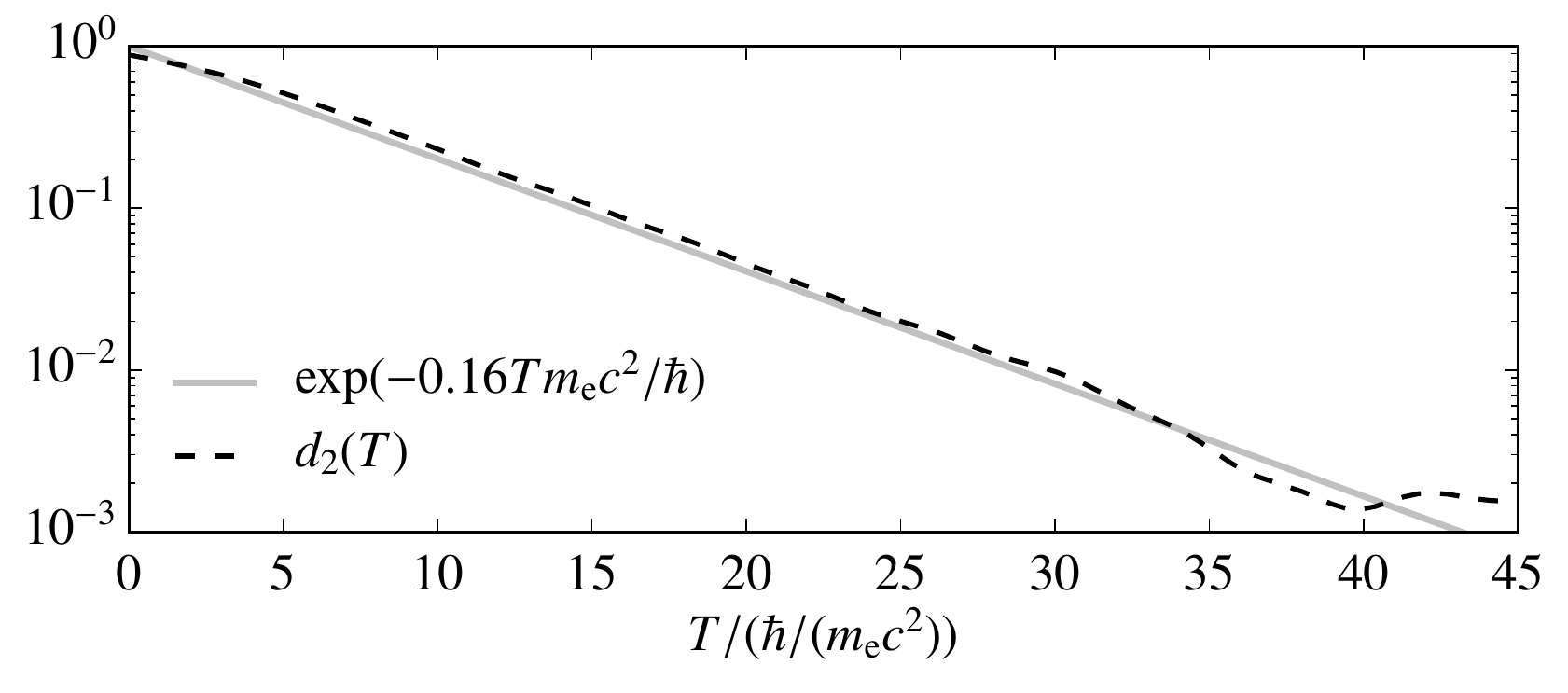}
  \caption{Decay probability $d_2$ as a function of the interaction time
    $T$ after the potential has been smoothly turned off (dashed black
    line) and an exponential fit to the data (solid gray line).
    The potential height is $V_0=-3.6m_\mathrm{e}c^2$ and other parameters
    are as in Fig.~\ref{fig:energy_spectrum_well}.}
  \label{fig:multi_pair_creation_d}
\end{figure}

The supercritical quasibound states are commonly interpreted as
different channels for pair creation.  As shown in
Refs.~\cite{Lv:Liu:etal:2013:Noncompeting_Channel,Lv:Liu:etal:2014:Degeneracies_of},
the complex scaling method can be utilized to determine the asymptotic
pair-creation rate for each channel.  As a (rather unexpected) central
result, there is no competition between the different pair-creation
channels, i.\,e., the number of created particles follows a
\emph{single} exponential law with a rate given by the sum of rates of
the individual channels.  Our results provide an intuitive explanation
of this phenomenon.  The presence of $n$ quasibound states favors the
population of multipair states; this means the \emph{simultaneous}
creation of $n$ particle-antiparticle pairs.  In other words, the $n$
supercritical quasibound states actually represent a single channel to
create number states with $n$ electron-positron pairs.  To provide a
more quantitative analysis, we define the quantity
\begin{equation}
  d_n(T) = | C_n(T\to\infty) - C_n(T) |\,.
\end{equation}
It characterizes how the initial vacuum state decays into a certain
number state that is occupied with probability \mbox{$C_n(T\to\infty)$}
for an asymptotically long interaction time~$T$.  The quantity $d_2(T)$
corresponding to the decay of the vacuum state into two-pair states is
shown on a logarithmic scale in Fig.~\ref{fig:multi_pair_creation_d}.
It follows approximately a straight line, which indicates that the decay
process is exponential, namely $d_2(T)\sim \exp(-\gamma T)$ with the
exponential parameter $\gamma\approx 0.16m_\mathrm{e}c^2/\hbar$.
According to Ref.~\cite{Lv:Liu:etal:2013:Noncompeting_Channel}, the
exponential decay rate of the occupation of the particle-number state to
its asymptotic value, which results from the complex scaling approach,
is $\gamma\approx 0.18m_\mathrm{e}c^2/\hbar$ for the same field
configuration as in Figs.~\ref{fig:multi_pair_creation_C_V0}(b)
and~\ref{fig:multi_pair_creation_d}.  The small discrepancy between
these two exponents may be attributed to the occurrence of the
intermediate single-particle states seen in
Fig.~\ref{fig:multi_pair_creation_C_V0}(b), which slows down the
transition from the vacuum into the final two-particle states.

\subsection{Position- and momentum-space distributions}

\begin{figure}
  \centering
  \includegraphics[scale=0.485]{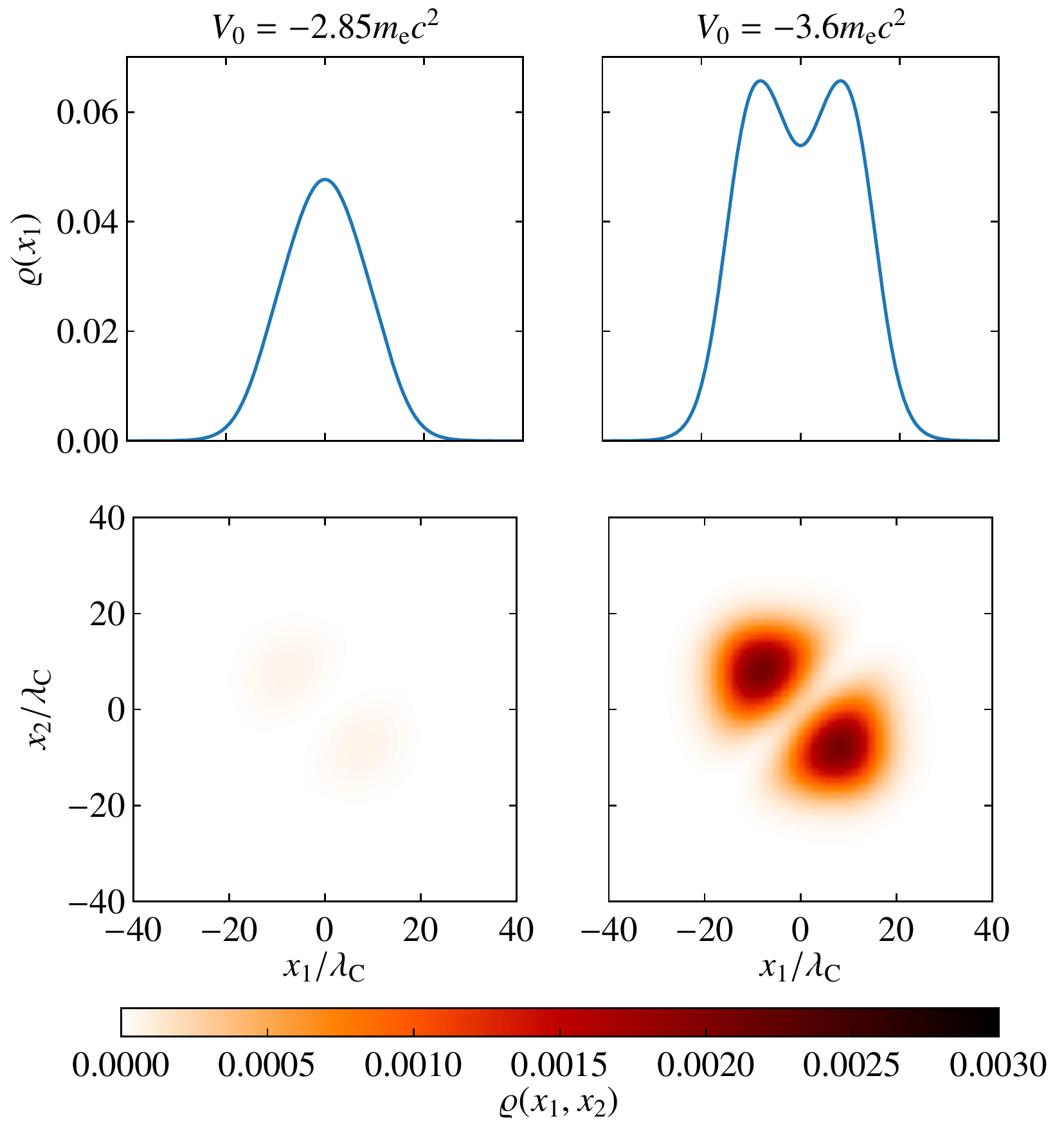}
  \caption{Single- and two-pair spatial densities $\varrho(x_1)$
    and $\varrho(x_1, x_2)$ as calculated for the parameters as in
    Fig.~\ref{fig:multi_pair_creation_C_V0} and a total interaction time
    $T=90\hbar/(m_\mathrm{e}c^2)$.  The left column corresponds to a
    potential strength of $V_0=-2.85m_\mathrm{e}c^2$ and the right one
    is for $V_0=-3.6m_\mathrm{e}c^2$.}
  \label{fig:multi_pair_creation_densities}
\end{figure}

Our numerical space- and time-resolved analysis of the pair-creation
dynamics allows us not only to determine how many particles and
antiparticles are created but also where these are created.
Figure~\ref{fig:multi_pair_creation_densities} shows the asymptotic
single- and two-particle probability densities $\varrho(x_1)$ and
$\varrho(x_1, x_2)$ for the position of the electrons as calculated for
the parameters as employed before and a potential strength of
$V_0=-2.85m_\mathrm{e}c^2$ and $V_0=-3.6m_\mathrm{e}c^2$, i.\,e., where
the final quantum field state consists asymptotically mainly of
single-pair states and two-pair states, respectively, as discussed
above.  The single-particle densities are concentrated in the vicinity
of the localized potential \eqref{eq:well_pot} for both cases.  However,
the origins of these distributions are different.  For the case
$V_0=-2.85m_\mathrm{e}c^2$, $\varrho(x_1)$ reflects the single-pair
states, which are the dominating states for this potential strength.
Furthermore, the two-particle density $\varrho(x_1, x_2)$ is close to
zero for $V_0=-2.85m_\mathrm{e}c^2$. For $V_0=-3.6m_\mathrm{e}c^2$ the
two-pair states are the dominating states and single-pair states are
not occupied as we mentioned before.  Consequently, the two-particle
density $\varrho(x_1, x_2)$ is substantial in this case.  It has two
maxima at $(x_1, x_2)\approx(-7\lambda_\mathrm{C}, 7\lambda_\mathrm{C})$
and $(x_1, x_2)\approx(7\lambda_\mathrm{C}, -7\lambda_\mathrm{C})$.  The
single-particle density $\varrho(x_1)$ is just the marginal distribution
of $\varrho(x_1, x_2)$ as there are no single-pair contributions to
$\varrho(x_1)$.  As a consequence of the fermionic nature of the created
particles and the Pauli exclusion principle, the density
$\varrho(x_1, x_2)$ vanishes along the line $x_1=x_2$.  Furthermore, the
two created electrons that emerge at $\pm 7\lambda_\mathrm{C}$ are
strongly correlated, i.\,e., the spatial distributions of the two
electrons are not statistically independent.

\begin{figure}
  \centering
  \includegraphics[scale=0.485]{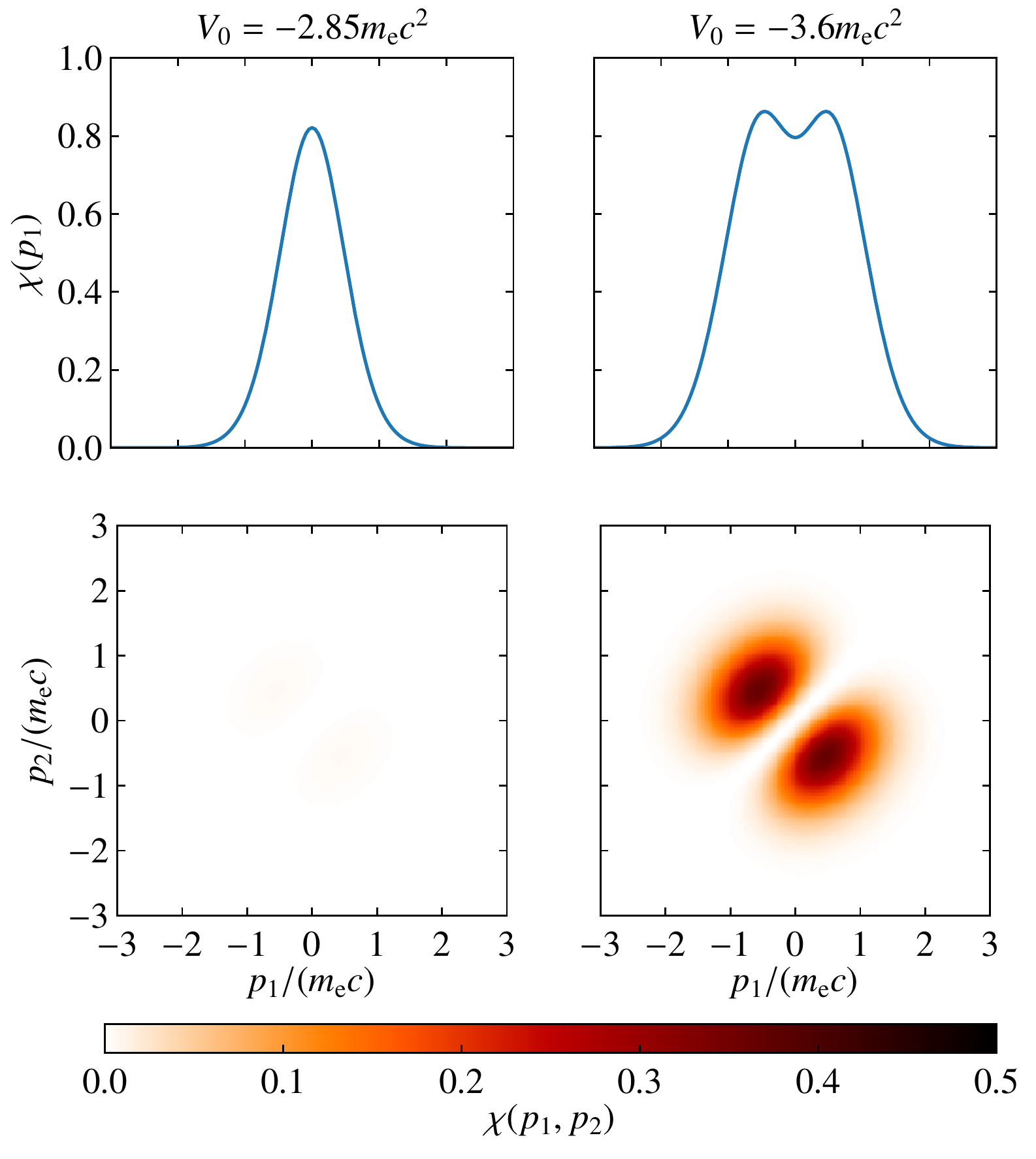}
  \caption{Single- and two-pair momentum spectra $\chi(p_1)$ and
    $\chi(p_1, p_2)$ of the created electrons as calculated for the
    parameters as in Fig.~\ref{fig:multi_pair_creation_C_V0} and a total
    interaction time $T=50\hbar/(m_\mathrm{e}c^2)$.  The left column
    corresponds to a potential strength of $V_0=-2.85m_\mathrm{e}c^2$
    and the right one is for $V_0=-3.6m_\mathrm{e}c^2$.}
  \label{fig:po_multi_pair_creation_p-spectrum_electron}
\end{figure}

\begin{figure}
  \centering
  \includegraphics[scale=0.485]{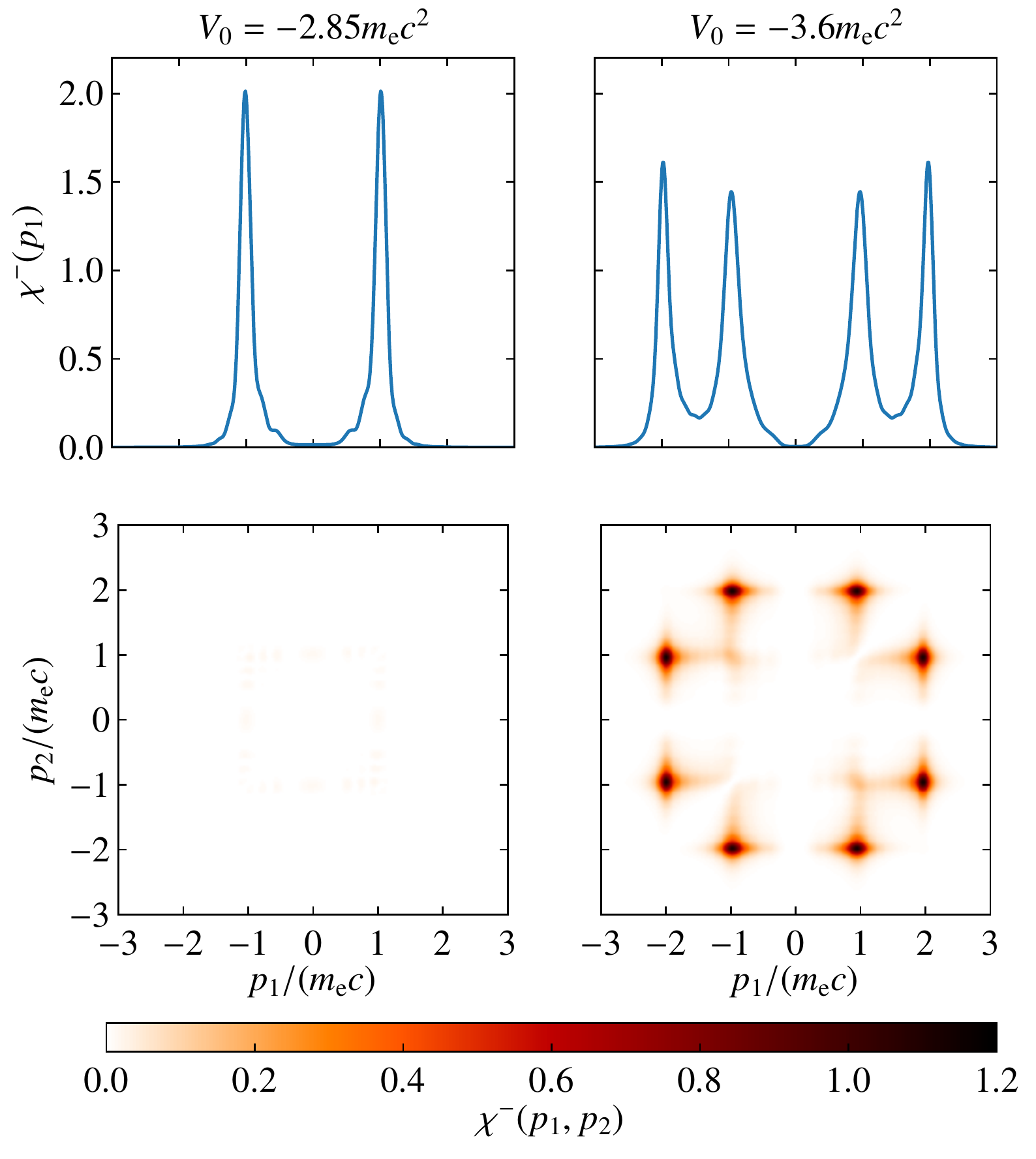}
  \caption{Single- and two-pair momentum spectra $\chi^-(p_1)$ and
    $\chi^-(p_1, p_2)$ of the created positrons as calculated for the
    parameters as in Fig.~\ref{fig:multi_pair_creation_C_V0} and a total
    interaction time $T=50\hbar/(m_\mathrm{e}c^2)$.  The left column
    corresponds to a potential strength of $V_0=-2.85m_\mathrm{e}c^2$
    and the right one is for $V_0=-3.6m_\mathrm{e}c^2$.}
  \label{fig:po_multi_pair_creation_p-spectrum}
\end{figure}

To complete our understanding of the decay process of the vacuum into
multiple electron-positron pairs, we also investigate the properties of
the created electrons and positrons in momentum space.  The positrons
are best characterized by its momentum spectrum, because for the
positrons the localized potential is repulsive leading to a strong
acceleration and delocalization.  In contrast, the electrons are
captured in the potential.  This means the momentum distribution of the
created electrons $\chi(p_1)$ is concentrated around zero as shown in
Fig.~\ref{fig:po_multi_pair_creation_p-spectrum_electron}.  Due to the
narrow potential and the resulting strong localization of the created
electrons, the electrons' momentum distribution is rather broad, i.\,e.,
it has a width of the order of $m_\mathrm{e}c$.  The momentum
distribution of the positrons $\chi^-(p_1)$ exhibits two peaks at
$p_1\approx\pm 1.01m_\mathrm{e}c$ for the parameter
$V_0=-2.85m_\mathrm{e}c^2$, as presented in
Fig.~\ref{fig:po_multi_pair_creation_p-spectrum}.  This means that the
individual positrons, which emerge for this parameter set, travel with
relativistic velocities to the left or to the right with equal
probability.  The distributions $\chi(p_1, p_2)$ and $\chi^-(p_1, p_2)$
almost vanish for $V_0=-2.85m_\mathrm{e}c^2$ as also indicated in
Figs.~\ref{fig:po_multi_pair_creation_p-spectrum_electron}
and~\ref{fig:po_multi_pair_creation_p-spectrum}, which again proves that
single-pair states dominate the pair-creation process in this case.  For
$V_0=-3.6m_\mathrm{e}c^2$, the creation process is triggered by two
different channels leading to an occupation of two-pair states and to a
rich structure of the asymptotic positron momentum distribution
$\chi^-(p_1, p_2)$.  This distribution features eight sharp maxima
approximately at $\pm0.96m_\mathrm{e}c$ and $\pm1.98m_\mathrm{e}c$.  As
there are no single-pair states for $V_0=-3.6m_\mathrm{e}c^2$, there are
no single-pair contributions to the momentum distributions
$\chi(p_1, t)$ and $\chi^-(p_1, t)$ which therefore follow from
$\chi(p_1, p_2)$ and $\chi^-(p_1, p_2)$, respectively, by integrating
out one momentum degree.  Because of the Pauli exclusion principle for
indistinguishable fermions, the corresponding two electrons/positrons
cannot have the same momentum and, therefore, the distributions
$\chi(p_1, p_2)$ and $\chi^-(p_1, p_2)$ vanish along the line $p_1=p_2$.

Signatures of the supercritical quasibound states are observable also in
the energy spectrum of the created electrons and positrons.  Most
likely, the created electrons have close-to-zero momentum, thus their
energy distribution is sharply peaked at $m_\mathrm{e}c^2$.
Furthermore, the energy of the created positrons equals approximately
the absolute value of the energy of the supercritical quasibound states.
For $V_0=-2.85m_\mathrm{e}c^2$, for example, the individual positrons of
the single-pair states travel with relativistic velocities to the left
or right with equal probability having a total energy of about
$1.4m_\mathrm{e}c^2$, which equals the absolute value of the energy of
the supercritical quasibound state for this potential strength; see
Fig.~\ref{fig:energy_spectrum_well}.  For $V_0=-3.6m_\mathrm{e}c^2$, the
energies that correspond to the positions of maxima in the momentum
distribution of Fig.~\ref{fig:po_multi_pair_creation_p-spectrum} are
$1.39m_\mathrm{e}c^2$ and $2.22m_\mathrm{e}c^2$, which agree again with
the energy values of the supercritical quasibound states in
Fig.~\ref{fig:energy_spectrum_well}.  Consequently, the net energy of
the created particles equals approximately the energy gap between the
energies of the supercritical quasibound states and the rest-mass energy
of a real particle, i.\,e., $m_\mathrm{e}c^2$.

\section{Conclusions}
\label{sec:Summa}

In this work, we applied a time- and space-resolved formulation of
quantum field theory to study fermionic strong-field pair creation.  The
theoretical foundation of this framework is based on the observation
that the quantum field theoretical state can be directly related to the
evolution of the time-dependent field operator in the Heisenberg
picture.  The evolution of the latter can be obtained from the dynamics
of single-particle states of the Dirac Hamiltonian.  This approach
provides access to the full quantum field states and to expectation
values of any physical observable for arbitrary external field
configuration, e.\,g., the probability to create a certain number of
electron-positron pairs.  The Coulomb binding energy can become
supercritical, i.\,e., quasibound states embedded in the negative-energy
continuum initiate pair creation from the vacuum.  For bosonic systems,
it was demonstrated in Ref.~\cite{Lv:Bauke:etal:2016:Bosonic_pair} that
the energy spectrum of the corresponding single-particle Hamiltonian
determines the dynamics of the pair-creation process.  Here, we studied
the relation between the energy spectrum and pair creation for fermionic
systems.

As a main result, we showed that the number of quasibound states of the
Dirac Hamiltonian at maximal potential strength equals the number of
created electron-positron pairs for a sufficiently long interaction
time.  This means that the pair-creation dynamics populates selectively
multipair states with a specific number of electron-positron pairs.
Consequently, one can control the number of created pairs by varying the
applied potential such that it has a specific number of supercritical
quasibound states.  Furthermore, the sum of the mean energy of the
created particles equals approximately the difference between the energy
of the quasibound states and $m_\mathrm{e}c^2$ summed over all
quasibound states.  As the applied localized potential is attractive for
electrons their energy is close to their rest-mass energy, while
positrons are accelerated to relativistic velocities.  If there are
several electrons/positrons created, they are strongly statistical
dependent as a consequence of the Pauli exclusion principle.

\acknowledgments

This work was supported by the Alexander von Humboldt Foundation.
Q.~Z.~Lv enjoyed several helpful discussions with Dr. Q.~Su,
Dr. R.~Grobe, Dr. M. Jiang and Dr. Y.~T.~Li at the onset of this work.

\bibliography{multi_pair_creation}

\end{document}